\documentclass[a4paper,12pt]{article}
\usepackage{epsfig}
\usepackage{amssymb}
\begin{document}
\thispagestyle{empty}

\newcommand{\etal}  {{\it{et al.}}}  
\def\Journal#1#2#3#4{{#1} {\bf #2}, #3 (#4)}
\def\PRD{Phys.\ Rev.\ D}
\def\NIMA{Nucl.\ Instrum.\ Methods A}
\def\PRL{Phys.\ Rev.\ Lett.\ }
\def\PLB{Phys.\ Lett.\ B}
\def\EPJ{Eur.\ Phys.\ J}
\def\IEEETNS{IEEE Trans.\ Nucl.\ Sci.\ }
\def\CPCD{Comput.\ Phys.\ Commun.\ }

\bigskip

{\bf
\begin{center}
 \textbf{\large {BEC of Two Photons and Higgs Physics} }
\end{center}
}


\begin{center}
{\large G.A. Kozlov  }
\end{center}


\begin{center}
\noindent
 {
 Bogolyubov Laboratory of Theoretical Physics\\
 Joint Institute for Nuclear Research,\\
 Joliot-Curie st., 6 Dubna\\
 141980 Moscow region, Russia
 }
\end{center}

\begin {abstract}
\noindent
{It is well understood that the studies of correlations between produced
particles, the effects of coherence and chaoticity, an estimation of
particle emitting source size play an important role in high energy
physics [1]. We mean the investigation of the space-time extension
or even squeezing of particle sources via the multiparticle quantum-statistics
correlation. We obtain the two-photon correlation function that
can provide the space-time information about the Higgs-boson source in thermal
environment and we argue that such an investigation could probe the Higgs-boson mass.}
\end {abstract}



\begin{center}
\noindent
\end{center}



1. The Large Hadron Collider (LHC) at CERN is at the stage to provide particle physicists with treasure of data.
These data allowed a precise measurement of many important parameters of modern particle physics 
in order to test their consistency
and to discover the Higgs-boson. At the same time, little is known about Higgs-gauge bosons interplay
in particular in theoretical aspects. The effect of Bose-Einstein correlations (BEC) is clear and undeniable
part of this theory, complicating the quantum statistical description of multi-lepton final states.

Historically, the BEC measurements have concentrated on pion pairs correlations, however have also been 
applied to more heavy hadrons, quarks, protons, and even gauge bosons, photons, $Z$-bosons etc.

It is known that the wave function describing a system of two identical bosons should be symmetric
under permutation of these bosons. As a consequence, the four-momentum differences between two bosons
will be smaller than in the world where Bose-Einstein statistics would not apply.

Actually, the direct photon BEC can provide information about the space-time distribution of the hot 
 matter prior to freeze-out. However, the BEC with direct (primary) photons is faced with difficulties 
 compared to hadron BEC primarily due to the small yield of photons emitted directly from the hot region 
 just after particle collisions. The main background of photons is produced by decays of final hadrons 
 or gauge bosons, or even (pseudo)scalar particles.

In Higgs searching program at the LHC, we suppose that two photons are produced mainly through Higgs-boson decay.
From the theoretical point of view the decay of Higgs into two photons emerges through the one-loop pattern:
the Higgs-boson transferred in pair of a quark and antiquark or vector bosons, or even in scalar particles.

In order to reject most of the hadron background, the reconstructed events in a spectrometer were required to have
a deposit energy (level) of greater than the energy well above the minimal energy at which photons emerge from hadrons.
Experimentally, the $\pi^{0}$ background constitutes a major difficulty which, however, to some extent,
can be taken care of by measuring the photon pair invariant mass.

The two-photon correlations in Higgs physics provide a powerful tool to explore 
the Higgs-boson mass estimated via the correlation radius which defines the geometrical size
of the two-photon  source.
Hence, the two-photon correlation function is strongly dependent on the space-time
properties of the Higgs decays.

In the quark-loop scheme, the non-relativistic bremsstrahlung formula for the current in quark-antiquark
interaction for one photon emission (with four-momentum $k = (k^{0},\,\vec {k})$) is
$$j^{\lambda}(k) =\frac{i\,e}{m_{q}\,k^{0}}\,\vec{p}\cdot\vec{\epsilon}_{\lambda}(k)\,e^{-k^{0}/\varepsilon_{0}}, $$
where $\vec{p}$ is the difference between the spatial momenta of a quark and antiquark with the mass $m_{q}$,
$\vec{\epsilon}_{\lambda}(k)$ is the vector of linear polarization of a photon; $\varepsilon_{0}$ means the
phenomenological parameter which depends on the initial energy.
For two sources the transition current is
$$ J^{\lambda}(k) =\sum_{n=1}^{2} e^{i\,k\,y_{n}}\,j^{\lambda}_{n} (k).$$
Index $n$ labels the independent quark-antiquark interactions
taking place at different space-time points $y_{n}$ which are considered to be
randomly distributed in the space-time volume of the (photon) source.
We suppose the gamma-quanta are created by some random currents (sources) $J_{\mu}(x)$
through the Lagrangian density $L_{int} = - g\,J_{\mu}(x)\,A^{\mu}(x)$. The neutral
weak current of (final) charged leptons is $J_{\mu}^{l}(x) = - \bar g\,\bar l(x)\,\gamma_{\mu}\,
l(x)$ with $l: e,\mu,\tau$, for which the problem can be solved exactly. Both the currents
$J_{\mu}(x)$ and $J_{\mu}^{l}(x)$ exist in a restricted space-time region and they are 
chaotically and randomly disturbed by external fields (forces). The restricted domain is
characterized by the internal stochastic scale $L_{st}$, the meaning of which is explained in [2,3].

By studying BEC of identical particles, it is possible
to determine the time scale and spatial region over which particles do not
have the interactions. Such a surface is called as decoupling one.
In fact, for an evolving system such as $p p$ collisions, it is
not really a surface, since at each time there is a spread out
surface due to fluctuations in the last interactions, and the shape
of this surface evolve even in time. The particle source is not
approximately constant because of energy-momentum conservation
constraint.

Actually, the second order distribution function 
$$ N_{12}(k_{1}, k_{2}) =
\langle\vert J^{\lambda_{1}}(k_{1})\,J^{\lambda_{2}}(k_{2})\vert^{2}\rangle$$
normalized to the product $N_{1}(k_{1})\cdot N_{2}(k_{2})$ with
one-particle distribution functions $N_{i}(k_{i}) =
\langle\vert J^{\lambda_{i}}(k_{i})\vert^{2}\rangle$ ($i=1,2$) formally
defines the probability to find two photons with momenta $k_{1}$ and $k_{2}$
issued at $y_{1}$ and $y_{2}$. The crossing momenta has to be taken into account.

In this work, we make an attempt to demonstrate that the problem
of properties of the genuine interactions can be explored using
experimental data which can be collected at the LHC.
These data can be analyzed through the compared
measures of some inclusive distributions and final state
correlations.

One of the aims of this paper is to carry out the  proposal for
the experimental measurements of virtual $\gamma^{\star}\gamma^{\star}$ pair correlations.

This exploration is theoretically supported by the quantum
field theory model approaches [4-9,2,3] at finite temperature, $({QFT}_\beta)$, where
one of the main parameters is the temperature of the particle (emitting) source under
the random external forces (fields) influence.

We propose that photons do not strongly interact with (produced) medium: they carry
information about early stage of reaction. Our pragmatic definition is: photons are
produced not from hadronic decays. Any source of real gamma-quanta produces virtual
photons with very low mass. If the momentum of the virtual gamma-quantum is sufficiently
small, the source strength should be small as well. The real gamma-quantum can be
measured from the virtual yield which is observed as low mass of the spectrum
for lepton-antilepton pair.

The main channels are the two-photon production
$pp\rightarrow Higgs\rightarrow\gamma^{\star}\gamma^{\star}\rightarrow
2\mu^{-}2\mu^{+},~2e^{-}2e^{+},~e^{-}e^{+}\mu^{-}\mu^{+},~...$ in $pp$ collisions.
An efficient selection of leptons needs to be  performed
according to the following criteria. First, all leptons were
required to lie in the pseudorapidity range covered by, e.g., the CMS
muon system that is, $|\eta| \le$ 2.4. Second, the leptons were
required to be unlikely charged in pairs.

The di-lepton channel is especially promising from the experimental
point of view, since it is expected that the experimental
facilities related for LHC will make it
possible to record muons of energy in the TeV range with a
resolution of about a few percent and an efficiency close to 100
\%. Moreover, this channel is characterized by a maximum
signal-to-background ratio in the energy region being considered.


2. A pair of identical bosons with momenta $p_{1}$ and $p_{2}$ and the
mass $m$ produced incoherently
from an extended source will
have an enhanced probability $C_{2}(p_{1},p_{2})=
N_{12}(p_{1},p_{2})/[N_{1}(p_{1})\cdot N_{2}(p_{2})]$ to be measured
in terms of differential cross section $\sigma$, where
 $$N_{12}(p_{1},p_{2})=\frac{1}{\sigma}\frac{d^{2}\sigma}{d\Omega_{1}\,d\Omega_{2}} $$
to be found close in 4-momentum space $\Re_{4}$ when detected
simultaneously, as compared to if they are detected separately with
$$ N_{i}(p_{i})=\frac{1}{\sigma}\frac{d\sigma}{d\Omega_{i}}, \,\,\,
 d\Omega_{i}=\frac{d^{3}\vec p_{i}}{(2\pi)^{3}\,2E_{p_{i}}}, \,\,
 E_{p_{i}}=\sqrt {\vec p_{i}^{2}+m^{2}},\,\,\,
 i = 1, 2. $$
In an experiment, one can account the inclusive density $\rho_{2}(p_{1}, p_{2})$
which describes the distribution of two particles in $\Omega$ (the sub-volume of the phase space)
irrespective of the presence of any other particles
$$ \rho_{2}(p_{1}, p_{2}) = \frac{1}{2!}\frac{1}{n_{events}}\,\frac{d^{2} n_{2}}{dp_{1}\,dp_{2}},$$
where $n_{2}$ is the number of particles counted in a phase space domain
$(p_{1} + dp_{1},p_{2} + dp_{2})$. The multiplicity $N$ normalizations stand as
$$\int_{\Omega} \rho(p)\,dp = \langle N\rangle, $$
$$\int_{\Omega} \rho_{2}(p_{1}, p_{2})\,dp_{1}\,dp_{2} = \langle N (N-\delta_{12})\rangle, $$
where $\langle N\rangle$ is the averaged number of produced particles. Here, $\delta_{12} =0$
for different particles, while $\delta_{12} =1$ in case of identical ones (coming from the same event).

On the other hand, the following relation can be used to retrieve the BEC function
$C_2(Q)$:
\begin{equation}
\label{e23}
 C_2(Q) = \frac{N(Q)}{N^{ref}(Q)},
\end{equation}
where $N(Q)$, in general case, is the number of particle pairs
(off-shell photons) in BEC pattern with 
\begin{equation}
\label{e24}
 Q = \sqrt {-(p_1-p_2)_{\mu}\cdot (p_1-p_2)^{\mu}}= \sqrt{M^{2} - 4\,m^{2}}.
\end{equation}
In definitions  (\ref{e23}) and  (\ref{e24}),  $N^{ref}$ is the number of
particle pairs without BEC and
$p_{\mu_{i}}= (\omega_{i}, \vec p_{i})$ are four-momenta of produced photons $(i = 1,\ 2)$;
$M = \sqrt {(p_1+p_2)^{2}_{\mu}}$ is the invariant mass of the pair of photons.

An essential problem in two-particle correlations is the estimation of the
reference distribution $N^{ref}(Q)$ in Eq. (\ref{e23}). If there are other
correlations beside the Bose-Einstein effect, the distribution $N^{ref}(Q)$ should be
replaced by a reference distribution corresponding to the two-particle distribution
in a geometry without BEC. Hence, the expression (\ref{e23}) represents the ratio
between the number of $\gamma^{\star}\gamma^{\star}$ pairs $N(Q)$ in the real world and the reference sample
$N^{ref}(Q)$ in the imaginary world. Note, that the reference sample can not be directly
observed in an experiment. Different methods are usually applied for the construction
of reference samples [1], however all of them have strong restrictions. One of the preferable
methods
is to construct the reference samples directly from data. 
The $\gamma^{\star}\gamma^{\star}$ BEC can be estimated for each bin of the photon average 
transverse momentum $p_{T} = \vert \vec {p}_{T_{1}} + \vec {p}_{T_{2}}\vert /2$ as the ratio 
of the distribution of photon pair invariant relative momenta where both photons with 
transverse momenta $\vec {p}_{T_{1}}$ and  $\vec {p}_{T_{2}}$ were taken from the same event to 
the same distribution but with the photons of the pairs taken from different events.
For our aim, for the reference sample
$N^{ref}(Q)$, it is suitable to use the pairs $\gamma^{\star}\gamma^{\star}$ from different (mixed) events.

It is commonly assumed that the maximum
of two-particle BEC function $C_2(Q)$ is 2 for $\vec p_{1} =
\vec p_{2}$ if no any distortion and final state interactions are
taking into account.


In general, the shape of $C_2(Q)$ is model dependent.
The most simple form of Goldhaber-like parameterization for $C_2(Q)$
[10,11] is often used for experimental data fitting.
\begin{equation}
 \label{e25}
 C_2(Q)=C_0\cdot (1+\lambda e^{-Q^2R^2})\cdot (1+\varepsilon Q) ,
\end{equation}
where $C_0$ is the normalization factor, $\lambda$ is the
chaoticity strength factor, meaning $\lambda =1$ for fully
incoherent and $\lambda =0$ for fully coherent sources; the
symbol $R$ is often called as the "correlation radius", and assumed to be
spherical in this parameterization.  The
linear term in (\ref{e25}) is supposed to be account within the
long-range correlations
outside the region of BEC. Note that the distribution of bosons
can be either far from isotropic, usually concentrated in some directions, or
almost isotropic, and what is important that in both cases the particles
are under the random chaotic interactions caused by other fields in the thermal
bath. In the parameterization (\ref{e25}) all of these issues
are embedded in the random chaoticity parameter $\lambda$.
To advocate the formula (\ref{e25}) it is assumed: \\
a. incoherent average over particle source where $\lambda$ serves to account for: \\
- partial coherence,\\
- $\gamma\gamma$ purity;\\
b. spherical Gaussian density of particle emission cell (with radius $R$);\\
c. static source which means no time (energy) dependence.\\
However, to enlarge the quantum pattern of particle production process and to avoid the static
and undistorted character of particle emitter source, we have already suggested to
use the $C_2(Q)$ function within $QFT_\beta$ accompanying by quantum evolution approach
in the form [2,3]:
\begin{equation}
\label{e26}
C_2(p_{1},p_{2}) \simeq \xi(N) \left \{1 + \lambda_{1}(\beta)\,e^{-\Delta_{p\Re}}
\left [1+\lambda_{2}(\beta)\,e^{ \Delta_{p\Re}/2}\right ]\right\} ,
\end{equation}
where $ \exp (-\Delta_{p\Re})$ is the smearing smooth dimensionless generalized function with 
$\Delta_{p\Re} =  (p_{1} - p_{2})^{\mu}\,\Re_{\mu\nu}\,(p_{1} - p_{2})^{\nu}$. $\Re_{\mu\nu}$ is the 
nonlocal structure tensor of the space-time size and it defines the domain of emitted photons. 
$\xi(N)$ depends on the multiplicity $N$ as $ \xi(N)= \langle{N (N-1)}\rangle/\langle N\rangle^2$.
The functions $ \lambda_{1} (\beta)$ and  $ \lambda_{2} (\beta)$ are the measures
of the strength of BEC between two photons: $\lambda_{1} (\beta)=
\gamma(\omega,\beta)/(1+\alpha)^{2}$, and the correction to the coherence
function in the brackets of Eq. (\ref{e26}) is
$\lambda_{2}(\beta) = 2\,\alpha/\sqrt{\gamma(\omega,\beta)}$.
 The function $\gamma (\omega,\beta)$ calls for the quantum thermal features of BEC
pattern and is defined as
\begin{equation}
\label{e27}
 \gamma (\omega,\beta) \equiv \gamma (n)  = \frac{{n^2 (\bar \omega )}}{{n(\omega )\ n(\omega
 ')}} ,\ \
 n(\omega ) \equiv  n(\omega ,\beta ) =
 \frac{1}{{e^{\omega \beta} - 1 }} ,\ \
 \bar\omega  = \frac{{\omega  + \omega '}}{2} , 
\end{equation}
where $n(\omega,\beta )$ is the mean value of quantum numbers for Bose-Einstein
statistics particles with the energy $\omega$ in the thermal bath
with statistical equilibrium at the temperature $T= 1/\beta$. The
following condition $\sum_{f} n_{f}(\omega,\beta) = N$ is evident,
where the discrete index $f$ stands for the one-particle state $f$.

The important parameter $\alpha (\beta)$ in (\ref{e26}), the measure of chaoticity,
summarizes our knowledge of other than space-time characteristics of the particle
emitting source, and it varies from $0$ to $\infty$ (see [12] for details).

In terms of time-like $R_{0}$, longitudinal $R_{L}$ and transverse $R_{T}$ components of the 
space-time size $R_{\mu}$, the distribution $\Delta_{p\Re}$ looks like 
\begin{equation}
\label{e266}
\Delta_{p\Re}\rightarrow \Delta_{pR} = (\Delta p^{0})^{2}\,R_{0}^{2} + 
(\Delta p^{L})^{2}\,R_{L}^{2} + (\Delta p^{T})^{2}\,R_{T}^{2}.
\end{equation}
$R_{0}$ in (\ref{e266}) is treated as the measure of the particle emission time, 
or even it represents the interaction strength of outgoing particles.

 Hence, we have introduce a new parameter $R_{\mu}$, a four-vector, which defines the region 
 of nonvanishing particle density with the space-time extension of the particle emission source. 
 Formula (\ref{e26}) must be understood in the sense that $ \exp (-\Delta_{p\Re})$ is a distribution that 
 in the limit $R\rightarrow\infty $ strictly becomes a $\delta$ - function. For practical using with 
 ignoring the energy-momentum dependence of $\alpha$, one has:
\begin{equation}
\label{e26666}
C_2(Q) \simeq \xi(N) \left \{1 + \lambda_{1}(\beta)\,e^{-Q^2 R^{2}}
\left [1+\lambda_{2}(\beta)\,e^{+Q^{2} R^{2}/2}\right ]\right\}. 
\end{equation}
The parameter $R$ in formula (\ref{e26666}) is the measure of the space-time
overlap between two photons, and the physical meaning of $R$ depends on the fitting of
$C_{2}(Q)$-function. 
$R$ can be defined through the evaluation of the root-mean-squared momentum $Q_{rms}$ as:
$$Q_{rms}^{2} (\beta) =\langle \vec Q^{2}\rangle =
\frac{\int_{0}^{\infty} d\vert\vec Q\vert\, \vec Q^{2}\,\left [\tilde C_{2}(Q,\beta) -1 \right ]}
{\int_{0}^{\infty} d\vert\vec Q\vert\,\left [\tilde C_{2}(Q,\beta) -1 \right ]},\,\,\,\,
\tilde C_{2}(Q,\beta) =\frac{C_{2}(Q,\beta)}{\xi (N)},$$
where $R$ and $Q_{rms}(\beta)$ are related to each other by means of  
$$R = R(\beta) = \left [\frac{3}{2}\left (1+\frac{1}{1+\frac{1}{4\,\alpha(\beta)}\,\sqrt \frac{{\gamma (n)}}{2}}\right )
\right ]^{1/2}\frac{1}{Q_{rms}(\beta)}. $$
We find the following restricted window $\sqrt {3/2} < (R\cdot Q_{rms}(\beta)) < \sqrt{3}$, where
the lower bound satisfies to the case $\alpha\rightarrow 0$ (no any distortion in the particle production
domain), while the upper limit is given by the very strong influence of chaotic external fields (forces),
$\alpha\rightarrow \infty$. The result is rather stable in the wide range of variation of $\alpha$.

3. It has been emphasized [2,3] that
there are two different scale parameters in the model considered
here. One of them is the so-called "correlation radius" $R$ introduced
in (\ref{e25}) and also presented in (\ref{e26}). In fact,
this $R$-parameter gives the pure size of the particle emission
source without the influence of the distortion and interaction forces coming from
other fields. The other (scale) parameter is the
scale $L_{st}$ of the production particle domain where the stochastic,
chaotic distortion due to environment (the influence of other fields, forces)
is enforced.
This stochastic scale carries  the dependence of the particle mass,
the $\alpha$-coherence degree and what is very important - the
temperature $T$-dependence.

One question arises: how can BEC be used to determine the effective scale $L_{st}$
and, perhaps, the phase transition?
We suppose the changes of $\gamma^{\star} \gamma^{\star}$ 
production region and effects
yielding the dynamical variables and parameters of BEC due to in-medium distortion.

Consider the finite system (expanding or squeezing) with the flow of two gauge bosons pairs,
eg., $B_{\mu}B_{\mu}$- pairs. The Hamiltonian is
$$H_{0} = \frac{1}{2}\int d^{3}x\,\left [\left (\dot{\phi}\right )^{2} +
{\vert\nabla \vec B\vert}^{2} + m^{2}\,B_{\mu}^{2}\right ]$$
which is asymptotically free in the rest frame of undistorted matter with the field
$B_{\mu} = (\phi,\vec B)$ having the mass $m$ in general case.
This Hamiltonian and commutation relations
can possess the exact symmetry. However, the observed states in real physical environment
can not be realized in the framework of this symmetry. $H_{0}$ has to be added
by the Hamiltonian
$$H_{dist} = \frac{g^{\mu\nu}}{2}\,\int d^{3}\vec{x}\,d^{3}\vec{y}\, B_{\mu}(\vec x)\delta F^{2}_{\beta}
(\vec x - \vec y)\,B_{\nu}(\vec y),$$
which is provided by the distortion due to in-medium effect, in particular, because of
temperature of the environment.
The field $B_{\mu}(\vec x)$ propagates in medium with $p$ (momentum) - dependent effective
(squeezing) frequency $\omega_{\beta} = \sqrt {m^{2} + \vec p^{2} -  \delta\tilde {F}^{2}_{\beta} (p)}$
and, consequently, the mass $m_{\beta}$ is related to the undistorted (asymptotic) mass $m$ by
$m_{\beta} = \sqrt {m^{2} - \delta\tilde {F}^{2}_{\beta} (p)}$.  Here, $\delta {F}^{2}_{\beta} (\vec x)$
is the non-local formfactor leading to the modification of the particle frequency (mass) of undistorted matter.
Because of the quadratic form of the Hamiltonian $H_{dist}$ through the asymptotic Bose - operators of
annihilation (creation) $a(p) (a^{+}(p))$ in 
$$B_{\mu}(x) =
\int \frac {d^{3}\vec {p}}{(2\,\pi)^{3}\,\sqrt {m^{2} + \vec {p}^{2}}}\,\sum_{\lambda}\,
\epsilon _{\mu}^{\lambda} (p)\,a(p)\,e^{-i\,p\,x}, $$ 
one can use the Bogolyubov  transformation
 $$b_{f} = u_{f}\,a_{f} - v_{f}\,a^{+}_{-f}, \,\,\,\,
 b_{f}^{+} = u_{f}^{\star}\,a_{f}^{+} - v_{f}^{\star}\,a_{-f} $$
 for new operators of annihilation (creation) $b_{f} (b^{+}_{f})$ in distorted medium.
 In some sense, the last operators correspond to thermalized-distorted quasiparticles.
  The functions $u_{f}$ and $v_{f}$ obey the condition ${\vert u_{f}\vert}^{2} -
 {\vert v_{f}\vert}^{2} =1$ and can be given in the form $u_{f} = \cosh \eta _{f}$,
 $v_{f} = \sinh \eta _{f}$, where $\eta_{f}$ has to be even function of index $f$.
 For  simplicity, we identify $f$ with momentum $p$,  and for squeezing frequency one gets
 $\eta_{p} = \ln \omega^{2} (p) / \omega^{2}_{\beta} (p)$. Indeed, the Bogolyubov
 transformation is equivalent to squeezing procedure.

4. In the Higgs-boson rest frame, there is a kinematical configuration
for two pairs of final lepton momenta $(p_{l}, p_{\bar l}^{\prime})$ and
$(q_{l}, q_{\bar l}^{\prime})$: $p_{l}\simeq p_{\bar l}^{\prime}$ and
$q_{l}\simeq q_{\bar l}^{\prime}$. Thus, the final state mimics a two-body final state,
and if the leptons being the electrons or even muons, the virtual photons with momenta $p$ and
$p^{\prime}$ are nearly on mass-shell, $p^{2}\simeq p^{{\prime}^{2}} \simeq 4\, m^{2}_{l}$,
where $m_{l}$ is the lepton mass. For the configuration given above,
the stochastic scale $L_{st}$ has different behavior depending on $T$.
At lower temperatures we have
\begin{equation}
\label{e31}
L_{st}\simeq {\left [\frac{\pi^{3/2}\, M_{H}^{2}\, e^{2\,m_{l}/T}}
{ 48\,\alpha(N)\, m^{11/2}_{l}\,T^{3/2}\, \left
(1+\frac{15}{16}\frac{T}{m_{l}}\right )}\right ]}^{\frac{1}{5}},
\end{equation}
where the condition $2\,n\,\beta\,m_{l} >1$  is taken into account for any integer $n$;
$M_{H}$ is the mass of the Higgs-boson, $m_{l}$ stands for the lepton mass.

On the other hand, at higher temperatures, when $T > 2\,n\,m_{l}$, one has
\begin{equation}
 \label{e32}
L_{st}\simeq {\left [\frac{\pi^{2}\, M_{H} ^{2}}{48\,\zeta (3)\, \alpha(N)\,
T^{3}\, m^{4}_{l}}\right ]}^{\frac{1}{5}},\,\,\,\,\,
\zeta (3) = \sum_{n=1}^{\infty} n^{-3} = 1.202.
\end{equation}

In case of two real photons correlation one can use the 
longitudinal stochastic scale instead of (\ref{e31}) and  (\ref{e32}), 
$L^{long}_{st} (m_{l} = m_{T}/2)$, with the average transverse mass 
$  m_{T} = 0.5 (\sqrt {p_{T_{1}}^{2}} + \sqrt {p_{T_{2}}^{2}})$ in 
the frame of, e.g., the Longitudinal Center of Mass System (LCMS) [13].

It turns out that the scale $L_{st}$ defines the range of stochastic forces.
This effect is given by $\alpha (N)$-coherence degree which can be estimated from
the experiment within the function $C_{2}(Q)$ as $Q$ close to zero, $C_{2}(0)$,
at fixed value of mean multiplicity $\langle N\rangle$:
\begin{equation}
 \label{e33}
\alpha (N) = \frac{1 + \gamma^{1/2}(n) -\tilde C_{2}(0) + \gamma^{1/4}(n) 
\sqrt {\tilde C_{2}(0) [\gamma^{1/2}(n) -2 ] + 2}}{\tilde C_{2}(0)-1},
\end{equation}
where $\gamma (n)$ is defined in (\ref{e27}) and $\tilde C_{2}(0)\equiv C_{2}(0) /\xi (N)$.
The upper limit of $C_{2}(0)$ depends on $\langle N\rangle$ and the quantum thermal features 
of BEC pattern given by $\gamma (n)$: $C_{2}(0) < \xi (N) [1 - \gamma^{1/2}(n)/2 ]^{-1}$. This 
upper limit is restricted by the maximal value of 2 in the ideal case as $\langle N\rangle \rightarrow\infty$ 
and $\gamma (n) =1$.


Note, that for $C_{2}(Q)$ - function (\ref{e26}), the limit $\alpha\rightarrow\infty$
yields for fully coherent sources with small $\langle N\rangle$, 
while $\alpha\rightarrow 0$ case stands for fully chaotic (incoherent) sources as 
$\langle N\rangle \rightarrow\infty$.
Actually, the increasing of $T$ leads to squeezing of the domain of
stochastic forces influence, and $L_{st}(T=T_{0})= R$ at some effective temperature $T_{0}$.
The higher temperatures, $T > T_{0}$,
satisfy to more squeezing effect and at the critical temperature
$T_{c}$ the scale $L_{st}(T=T_{c})$ takes its minimal value.
Obviously, $T_{c}\sim O(200~GeV)$ defines the phase transition where
the chiral symmetry restoration will occur.
Since in this phase all the masses tend to zero  and $\alpha\rightarrow 0$ at $T>T_{c}$, one
should expect the sharp expansion of the region with
$L_{st}(T>T_{c})\rightarrow \infty$.


Using the relation between $R_{L}$ and $m_{T}$ obtained from the Heisenberg uncertainty relations [14] 
$R_{L}(m_{T}) = c\sqrt{h\,R_{0}/2\,\pi\,m_{T}}$ one can estimate $R_{0}$ within the formula:
\begin{equation}
\label{e34}
R_{0}\simeq \frac{2\,\pi}{c^{2}\,h}{\left [\frac{\pi^{2}\,M_{H}^{2}}
{3\,\zeta (3)\, \alpha(N)\,T_{0}^{3}\, m_{T}^{4}}\right ]}^{\frac{2}{5}},
\end{equation}
where $c$ and $h$ are the speed of light in vacuum and  
the Planck constant, respectively. The expression (\ref{e34}) relates 
the two-photon emission time $R_{0}$ with the Higgs-boson mass $M_{H}$, the transverse mass 
of two photons $m_{T}$ and the temperature of the BEC domain $T_{0} > n\,m_{T}$. 
From  (\ref{e34})  one obtains $R_{0}\simeq 10^{-24} sec$ and $R_{0}\simeq 0.45\cdot 10^{-24} sec$ for 
$T_{0} = 50$ GeV and $T_{0} = 100$ GeV, respectively, at $M_{H} = 120$ GeV and $m_{T} = 1$ GeV 
in the case of two final muon pairs.
The confirmation of this should come from the measurement of the Higgs-boson lifetime in 
two-photon decay channel, as well as from the BEC data of electroweak interacting particles.

The qualitative relation between $R$ and $L_{st}$ above mentioned is the only one we
can emphasize in order to explain the mass and temperature dependencies of the source size. 
The dependence of the stochastic measure of chaoticity $\alpha$ on the minimum scale cut $L_{st}$ 
and $Q_{rms}$ can be used to define the fit region for different $p_{T}$. Such a minimum 
cut on $L_{st}$ introduces a lower cutoff on $R$.

There are  a number of effects which may give rise to $Q\simeq 0$ correlations and, thus, 
mimic the two-photons BEC coming from Higgs decays. These include $i)$ correlations of light 
hadrons and/or vector bosons misidentified as photons, $ii)$ radiative decays of resonances 
in both pseudoscalar and vector sectors, $iii)$ collective flow, etc.

Apparatus or analysis effects which may result in the sensitivity of external random forces 
influence, may be investigated by studying the dependence of the correlation functions on 
the stochastic scale $L_{st}$. This effect is expected to contribute strongly at small $L_{st}$ 
(or large $\alpha$ and $T$).  

Our model is consistent with the idea of the unification of weak and electromagnetic
interactions at $T > T_{c}$, predicted by Kirzhnitz and  Linde in 1972 [15,16].
In addition, there is the analogy with the asymptotic free theory:
the properties of environment (media) are the same as those  composed of free particles
in the infinite volume (Universe).



5. \textbf{To summarize:} one of the main reasons to study BEC at finite temperature
is the possibility to determine the precision with which the source size
parameter and the strength chaoticity parameter(s) can be measured at particle colliders.
Such investigations provide an opportunity
for probing the temperature of the particle production source and the
details of the external forces chaotic influence. Moreover, one can predict the mass
of the Higgs-boson. No systematic theoretical treatment of the Bose-Einstein effect within the Higgs decays
in  $\gamma^{\star}$ - pair production has been given so far. 

In this paper, we faced to the model $C_{2}(Q,\beta)$ - function in which the
contribution of $N$, $T$, $\alpha$ are presented.
This differs from the methods used in, e.g., LEP experiments based on the
approach (\ref{e25}). In fact, the latter resembles the traditional way of BEC study,
however any qualitative interpretations of $R$, $\lambda$, $\epsilon$ have no been  clarified.
The model proposed in this paper is
expected to be sensitive to the temperature of the environment and to the
external distortion effects.



We find that the stochastic scale $L_{st}$ decreases with increasing $T$ slowly at low
temperatures, and it decreases rather abruptly when the critical temperature is approached.

We obtain the dependence of the correlation strength functions $\lambda_{1}$ and $\lambda_{2}$ 
on the distance $L_{st}$ and the maximal value of $C_{2}$ at minimum invariant momentum $Q$ 
and $T$. 

We predict for the first time the spatial size of the source (the correlation radius $R$)
of two  photons originated from the Higgs-boson decay in restricted domain at the
proper temperature $T_{0}$. The dependence of the Higgs-boson mass, the mean multiplicity
$\langle N\rangle$ and the lepton mass is obtained in the form
$$ R \sim \frac{M^{2/5}_{H}\,e^{2\,m_{l}/{5T_{0}}}}{\alpha ^{1/5}(N)\, 
T_{0}^{3/10}\,m_{l}^{11/10}} $$
for low values of $T_{0} < n\,m_{l}$, while for higher temperatures one has
$$ R \sim \frac{M^{2/5}_{H}}{\alpha ^{1/5}(N)\, T_{0}^{3/5}\,m_{l}^{4/5}}. $$
 The  correlation radius $R$ increases with heavier Higgs-bosons at large gamma-quantum mean 
 multiplicity $\langle N\rangle$.
Actually, the experimental measuring of $R$ (in $fm$) can provide the precise estimation
of the effective temperature $T_{0}$ which is one of the main thermal characters
in the $\gamma ^{\star}\gamma ^{\star}$ pair emitter source
in the proper leptonic decaying channel $\gamma ^{\star}\gamma ^{\star}\rightarrow l\bar ll\bar l$
with the final lepton energy $\sqrt{\vec k^{2}_{l} + m^{2}_{l}}$ at
given $\alpha (N)$ fixed by $C_{2}(Q=0)$ and $\langle N\rangle$. $T_{0}$ is
the true temperature in the region of multiparticle production with dimension
$R = L_{st}$, because at this temperature it is exactly the creation of two
$\gamma$ quanta occurred in decay of Higgs-boson, and these particles obey the criterion of BEC.

The parameter $\alpha$ can be extracted from the experimental data on the two-photon
$\gamma^{\star}\gamma^{\star}$
BEC using $C_{2}(Q)$ with $Q$ being close to zero (see (\ref{e33})).
This allows one to estimate the Higgs-boson mass, $M_{H}$, at the temperature $T_{0}$.
As a qualitative illustration we present here an estimation of $M_{H}$ assuming
$T_{0} = 50, 100$ GeV, $\alpha = 10$ $\%$  
and $m_{l} = m_{\mu} = 105$ MeV. The result is rather sensitive
to the correlation radius $R$ and $T_{0}$: $M_{H}$ = 167 GeV at $T_{0} = 50$ GeV, 
$M_{H}$ = 477 GeV at $T_{0} = 100$ GeV for $R$ = 1.0 $fm$; $M_{H}$ = 68 GeV at $T_{0} = 50$ GeV, 
$M_{H}$ = 194 GeV at $T_{0} = 100$ GeV for $R$ = 0.7 $fm$;

Because of the fact that $(\mu^{+}\mu^{-})$ pairs originated from $\gamma^{\star}$ decay
overlap in space and are created in time almost simultaneously, it is natural to expect
that there are correlation between $(\mu^{+}\mu^{-})$ pairs coming from different
$\gamma^{\star}$'s due to Bose-Einstein interference.
These effects may also affect the accuracy with which the
$(\mu^{+}\mu^{-})$ pair mass can be measured at the LHC.
{
\begin{center}
REFERENCES

\end{center}

1. R.M. Weiner, Phys. Rep. 327 (2000) 249.\\


 2. G.A. Kozlov, Phys. Nucl. Part. Lett. 6 (2009) 162.\\

 3. G.A. Kozlov, Phys. Nucl. Part. Lett. 6 (2009) 177.\\

 4. G.A. Kozlov, Phys. Rev. C58 (1998) 1188.\\

 5. G.A. Kozlov, J. Math.  Phys. 42 (2001) 4749.\\

 6. G.A. Kozlov, New J. of Physics 4 (2002) 23.\\

 7. G.A. Kozlov, O.V. Utyuzh and G. Wilk, Phys. Rev. C68 (2003) 024901.\\

 8. G.A. Kozlov, J. Elem. Part. Phys. Atom. Nucl. 36 (2005) 108.\\
 
 9. G.A. Kozlov, O.V. Utyuzh, G. Wilk, W. Wlodarczyk, Phys. of Atomic Nucl.
  71 (2008) 1502.\\



10. G. Goldhaber et al., Phys. Rev. Lett. 3 (1959) 181.\\

11. G. Goldhaber et al., Phys. Rev. 120 (1960) 300. \\


12. G.A. Kozlov,  hep-ph/0512184.\\

13. T. Csorgo, S. Pratt, Proc. Workshop on Rel. Heavy Ion Phys. (Eds.
T. Csorgo et al.) 1991 75.\\

14. G.Alexander, Phys. Lett. B506 (2001) 45; 
G. Alexander and E.Reinhertz-Aronis, arXiv:0910.0138 [hep-ph].\\

15. D.A. Kirzhnits, JETP Lett. 15 (1972) 529.\\

16. D.A. Kirzhnits, A.D. Linde, Phys. Lett. 42B (1972), 471.\\






\end{document}